\documentclass[prl,aps,twocolumn,superscriptaddress,amsmath,amssymb,floatfix]{revtex4}
\pdfoutput=1
\usepackage{booktabs,graphicx,pstricks,verbatim}
\usepackage[colorlinks=true,citecolor=blue,filecolor=blue,linkcolor=blue,urlcolor=blue,pdftex]{hyperref}
\newcommand{\1}[1]{\, \mathrm{#1}} 
\newcommand{\n}[1]{\mathrm{#1}}    
\newcommand{\be}{\begin{eqnarray}}
\newcommand{\ee}{\end{eqnarray}}
\newcommand{\lsim}{\stackrel{<}{\sim}}
\newcommand{\gsim}{\stackrel{>}{\sim}}
\renewcommand{\section}[1]{\paragraph{#1:}}

\begin{document}

\title{Impure Thoughts on Inelastic Dark Matter}

\author{Spencer Chang}\affiliation{Physics Department, University of California Davis, Davis, CA 95616}\affiliation{Department of Physics, University of Oregon, Eugene, OR 974034}
\author{Rafael F. Lang}\affiliation{Physics Department, Columbia University, New York, NY 10027}
\author{Neal Weiner}\affiliation{Center for Cosmology and Particle Physics, Department of Physics, New York University, New York, NY 10003}


\begin{abstract}
The inelastic dark matter scenario was proposed to reconcile the DAMA annual modulation with null results from other experiments. In this scenario, WIMPs scatter into an excited state, split from the ground state by an energy $\delta$ comparable to the available kinetic energy of a Galactic WIMP. We note that for large splittings $\delta$, the dominant scattering at DAMA can occur off of thallium nuclei, with A$\sim$205, which are present as a dopant at the $10^{-3}$ level in NaI(Tl) crystals. For a WIMP mass $m_{\chi} \approx 100\1{GeV/c^2}$ and $\delta \approx 200\1{keV}$, we find a region in $\delta-m_{\chi}-$parameter space which is consistent with all experiments. These parameters in particular can be probed in experiments with thallium in their targets, such as KIMS, but are inaccessible to lighter target experiments. Depending on the tail of the WIMP velocity distribution, a highly modulated signal may or may not appear at CRESST-II.
\end{abstract}

\maketitle

\section{Introduction}
For more than a decade, the DAMA collaboration has employed ultra-pure NaI(Tl) crystals to search for dark matter scattering off a laboratory target. Their observation of a modulation in the spectral rate at low energies~\cite{Bernabei:2010mq} is a challenge to understand. While no obvious background can mimic this modulation, conventional models of dark matter explaining this modulation predict signals which would have long since been seen at other direct detection experiments~\cite{Aprile:2010um,Ahmed:2009zw,Angle:2007uj,Angloher:2008jj,Lee.:2007qn,Lebedenko:2008gb,Behnke:2008zza}. In light of these tensions, various proposals have been put forward to explain the DAMA modulation, such as light dark matter with or without ion channeling~\cite{Bottino:2003cz,Belli:1999nz,Petriello:2008jj,Chang:2008xa,Savage:2008er,Hooper:2010uy}, spin-dependent scattering~\cite{Ullio:2000bv,Belli:2002yt,Savage:2004fn,Fairbairn:2008gz}, mirror dark matter~\cite{Foot:2008nw}, momentum-dependent scattering~\cite{Feldstein:2009tr,Chang:2009yt} and inelastic dark matter (iDM)~\cite{TuckerSmith:2001hy,TuckerSmith:2004jv}.

In the iDM framework, WIMPs (weakly interacting massive particles) with mass~$m_{\chi}$ scatter only by transitioning to a heavier WIMP state~$\chi^*$, with mass splitting $\delta \equiv m_{\chi^*}-m_\chi \sim \mu v^2$ comparable to the available kinetic energy, which depends on the reduced mass~$\mu$. Among other things, this kinematical change pushes the expected signal to higher energy, increases the modulation amplitude, and favors heavier target materials. Together, these features allow a positive signal at DAMA while suppressing or eliminating signals at other experiments~\cite{Chang:2008gd,MarchRussell:2008dy,Kuhlen:2009vh,Kopp:2009qt}. However, recent results have placed iDM under increasing pressure. The null results at CRESST-II~\cite{Angloher:2008jj,SchmidtHoberg:2009gn}, CDMS~\cite{Ahmed:2009zw} and ZEPLIN-III~\cite{Akimov:2010vk} limit the allowed parameter space to non-Maxwellian halos~\cite{Lang:2010cd,Alves:2010pt}. While such properties may be natural, it is clear that iDM is now very constrained. As such, it is worth revisiting the original proposal to consider whether important effects may have been neglected.

\section{Inelastic Sensitivities}
Owing to the introduction of the splitting parameter~$\delta$, the kinematical requirement for scattering becomes
\be
v_{\chi,\n{min}} &=& \sqrt{\frac{1}{2 m_N E_R}} \left( \frac{m_N E_R}{\mu} + \delta \right),\label{eq:betamin}
\ee
where~$m_N$ is the mass of the target nucleus, $\mu$~the reduced mass of the WIMP-nucleus system and~$E_R$ the recoil energy. Due to this constraint, different target nuclei sample significantly different parts of the WIMP velocity distribution. Since the Maxwellian velocity distribution is falling exponentially at its tail, different targets with different threshold velocities~$v_{\chi,\n{min}}$ can have dramatically different sensitivities to a WIMP with given mass~$m_{\chi}$ and splitting~$\delta$. As a consequence, even sub-dominant components of the target can be the {\em dominant} source of scattering. In particular, for large values of~$\delta$, it may be impossible for iodine scatterings to occur at DAMA at all. However, while iodine is the heaviest element present at DAMA in large quantities, the thallium dopant is present at the~$10^{-3}$ level in the NaI(Tl) scintillator~\cite{Bernabei:2008yh}.

\section{Thallium as a Target}
Let us quantify whether the dominant scattering in DAMA may be arising from scattering off of the thallium which is present in the crystal. The tightest constraint on this high-$\delta$ region of iDM parameter space comes from CRESST-II. The atomic mass of tungsten ($A\sim184$) is nearly as large as that of thallium ($A\sim205$). Given the small concentration of thallium in DAMA, the observed modulation rate of $\sim0.04$~counts/day/kg of NaI(Tl) would correspond to an unmodulated event rate of $\sim40$~counts/day/kg if thallium were the target nucleus. As a consequence, the constraints from CRESST-II are that thallium can only be the target if scattering off of tungsten is nearly kinematically forbidden~\footnote{If the dark matter scatters inelastically with spin-dependent couplings~\cite{Kopp:2009qt}, this CRESST-II constraint goes away, since tungsten has almost no spin coupling, while thallium has a net proton spin. These scenarios would only have signals off of thallium doped scintillators. However, the cross sections required by the DAMA modulation are quite large, $\gsim 10^{-30}\; \n{cm}^2$, which are difficult to realize in particle physics models. Inelastic magnetic dipole scattering would be another possibility~\cite{Chang:2010en}, as thallium also has a large magnetic dipole, but again, the magnetic moment would have to be roughly larger than that of the proton.}.

\begin{figure*}[ht]
\hskip -0.1in \includegraphics[width=1.\columnwidth]{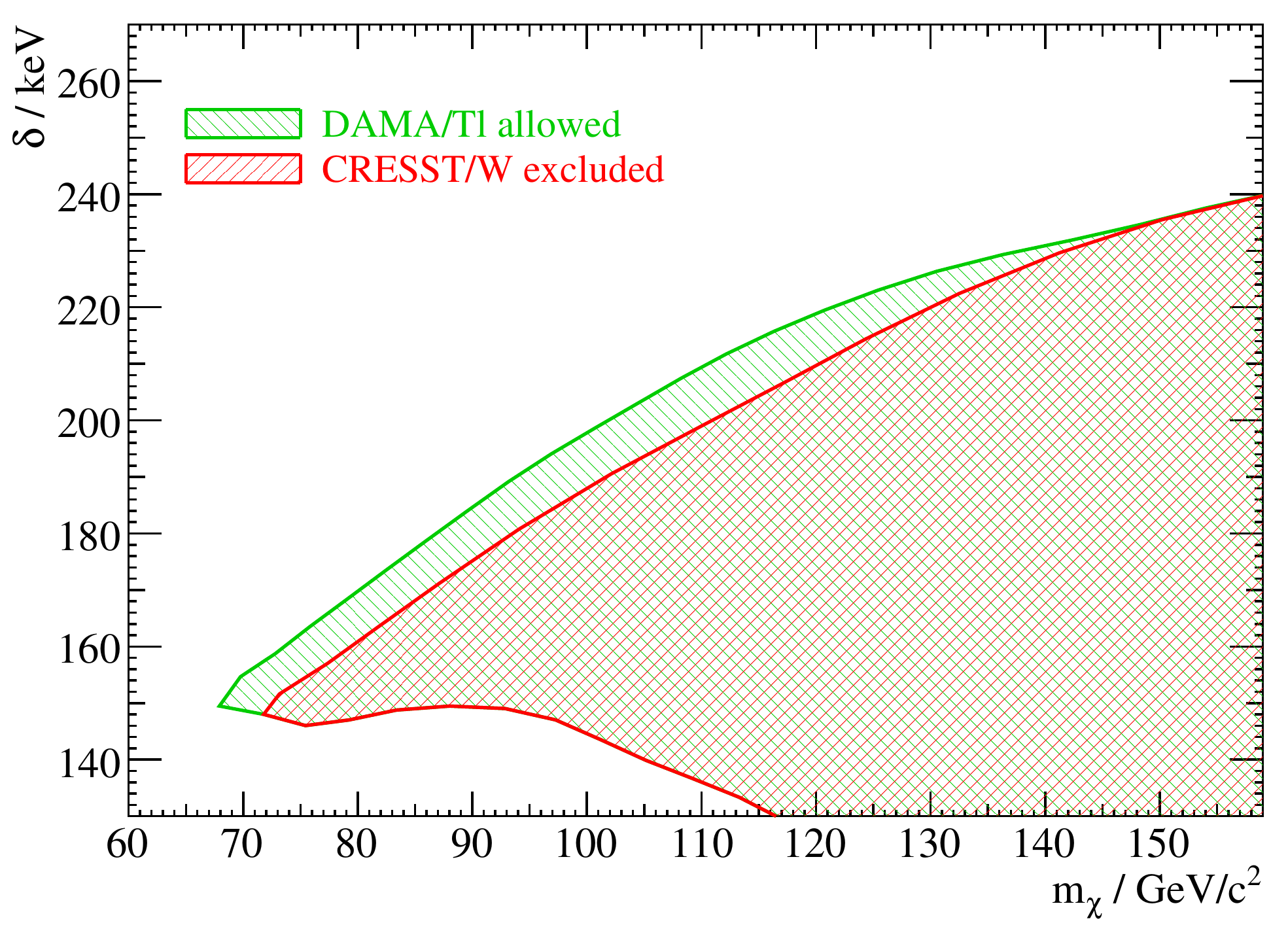}\hskip 0.2in \includegraphics[width=1.\columnwidth]{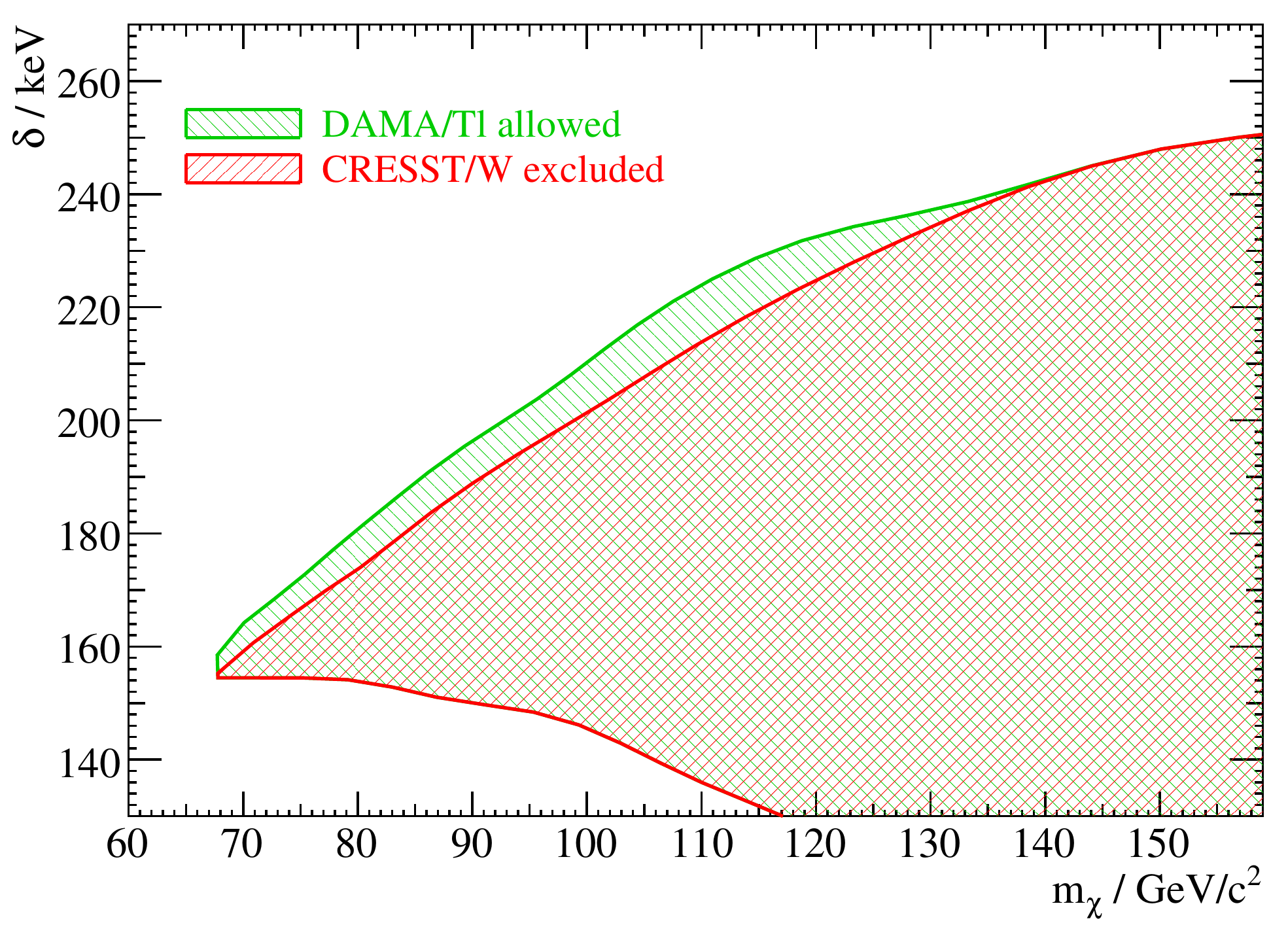}
\caption{The DAMA-allowed range of $\delta-m_{\chi}-$parameter space at 90\% confidence for $\chi-\n{Tl}$~scattering only (outer green hatched region) and constraints from CRESST-II (inner red hatched region). In the remaining allowed range of splittings~$\delta$, no scattering on sodium or iodine occurs. These contributions to signal at lower~$\delta$ are neglected here. {\em Left:} shown for a Maxwellian halo with $v_{\n{esc}}= 500 {\rm km/s}$, {\em right:} using the Via Lactea II simulation data (see~\cite{Kuhlen:2009vh}).}
\label{fig:deltamass}
\end{figure*}

We show the allowed iDM parameter space from DAMA including scattering off of thallium, together with the constraints from the tungsten target of CRESST-II in figure~\ref{fig:deltamass}. To this end, we calculate the annual modulation at each point in $\delta-m_{\chi}-$parameter space and compare it to the observed modulation in the binned $2-8\1{keV_{\n{ee}}}$ region from the recent DAMA data~\cite{Bernabei:2010mq}.

We use the Helm form factor~\cite{Helm:1956zz}, with parameterization as described in~\cite{Chang:2008xa}, and a Maxwellian velocity distribution with $v_0 = 220\1{km/s}$ and $v_{\n{esc}}=500\1{km/s}$. We note that our results do not qualitatively depend on this choice: for example, using $v_{\n{esc}}=600\1{km/s}$, the region looks similar but is shifted by about $40\1{keV}$ towards higher $\delta$. We further test the sensitivities by exploring other halo models. In particular, we use the tabulated Via Lactea II data from~\cite{Kuhlen:2009vh}, which leads to qualitatively similar results, as shown in figure~\ref{fig:deltamass}. We find the best $\chi^2$ fit parameter point for DAMA in cross section $\sigma_n$, splitting $\delta$, and mass $m_{\chi}$, and consider the 90\%~confidence region (or $\Delta \chi^2 < 6.25$) for these three parameters. This region, projected onto the $(\delta,m_{\chi})$ plane, is given by the green contour in figure~\ref{fig:deltamass}. For a given pair of $(\delta,m_{\chi})$, we choose the lowest cross section $\sigma_n$ consistent with the DAMA data at 90\% confidence, and evaluate the expected CRESST-II signal given this cross section. The cross sections per nucleon near the top of the green contour are $\sim 2 \times 10^{-34}\; \n{cm}^2$ and higher. While these large cross sections are difficult (if not impossible) to explain with standard-model mediators, they can be achieved with light mediators~\cite{ArkaniHamed:2008qn,Baumgart:2009tn,Cheung:2009qd,Morrissey:2009ur,Bjorken:2009mm}. To place limits, shown in red in figure~\ref{fig:deltamass}, we use the CRESST-II commissioning run release~\cite{Angloher:2008jj}, and require the signal be less than the 7 events observed. As one moves upward in $\delta$ in the allowed range of parameters, while keeping the DAMA modulation fixed, the event rate at CRESST-II is dropping rapidly. Consequently, one could employ more complicated techniques, such as the maximum gap technique, but these would all achieve essentially the same result, since over much of the allowed range of parameter space to explain DAMA, the scattering signal at CRESST is considerably less than 7 events.

\section{Discussion}
There are no measurements of the quenching factor for thallium, and thus we must make use of estimates. For the DAMA-allowed region, we assume the quenching $q$ is approximately proportional to the path length of the thallium nucleus in the crystal~\cite{Bavykina:2007ze}. We calculate this path length using the SRIM code~\cite{srim} and find that it scales approximately as $m_N^{-1}$. Thus, we conservatively take $0.88 > q_\n{Tl}/q_\n{I}> 127/205$, where the upper limit comes from the ratio of path-lengths using the SRIM database, and the lower comes from assuming a simple inverse proportionality to mass. Taking a range of $0.06<q_{\n{I}}<0.09$~\cite{Fushimi:1993nq,Tovey:1998ex,Gerbier:1998dm}, we find a range of quenching factors $ 0.037 \lsim q_{\n{Tl}} \lsim 0.08$, and conservatively combine all allowed regions in figure~\ref{fig:deltamass}. Given the uncertainties, we opt to include a broader range of quenching factors. Quantitatively, our fits prefer lower quenching factors ($q_{\n{Tl}} \lsim 0.05$), but this is a bit misleading. The dominant factor contributing to the fit is the location of the form factor zero of thallium. For $q_{\n{Tl}}\approx 0.07$, this happens to fall precisely in the middle of the DAMA energy range, using our Helm parameterization. However, this form factor has never been measured, and if its zero is shifted up even by 5\% in momentum transfer $q=\sqrt{2 m_{\n{Tl}} E_R}$, we can find good fits even with $q_{\n{Tl}}=0.08$. Given these uncertainties, we emphasize that only the specifically allowed range of parameters depends on the quenching factor, but not the presence of an allowed range itself.

The CaWO$_4$ crystals employed by CRESST-II contain impurities~\cite{Lang:2009ge}, for instance, Nd ($\lsim$ 1000 ppb), Gd ($\lsim$ 4000 ppb) and Sm/Dy/Hf/Os ($\lsim$ 20 ppb each), but in addition to their small abundances, these impurities are all too light to be relevant as a target. Heavier impurities are also irrelevant due to their small abundances. Even taking a high rate of $\sim 600\1{\mu Bq}$ from $^{210}\n{Pb}$~\cite{Lang:2009wb}, the concentration of $^{210}\n{Pb}$ compared to $^{\n{nat}}\n{W}$ is $\sim 10^{-18}$. Concentrations of $^{235}\n{U}$, $^{238}\n{U}$ and $^{232}\n{Th}$~\cite{Cozzini:2004vd} can be estimated to be below $10^{-11}$ compared to $^{\n{nat}}\n{W}$, and thus are safely sub-dominant.

The KIMS experiment uses doped CsI(Tl) targets~\cite{Lee.:2007qn}, also with a thallium concentration at the $10^{-3}$ level~\cite{Park:2002jr}. However, the event rate $\sim 0.28 \pm 0.18$~counts/day/kg~\cite{Chang:2008gd} observed in KIMS is consistent with a non-zero value. Since CRESST-II pushes us to large values of~$\delta$, the modulation should be~$\mathcal{O}(1)$~\cite{Chang:2008gd}, and the DAMA modulation rate of 0.04~counts/day/kg should be interpreted as roughly the average annual rate, as well. Thus, KIMS does not place a strong constraint. However, it is important to note that KIMS should necessarily see the modulation, given the doping concentration is similar to that of DAMA.

We have performed the analysis using a Maxwellian approximation, but this is well-known to be a poor approximation for iDM~\cite{Kuhlen:2009vh}. Consequently, we must be cautious in our predictions for CRESST-II. Clearly, the signal at CRESST-II must be highly suppressed to compensate for the small concentration of thallium at DAMA. Within the Maxwellian approximation, it appears that the event rate is naturally at higher ($\gsim 40 \1{keV}$) energies. Nonetheless, the presence of high-velocity structures could be relevant and cause a non-negligible signal to appear at CRESST-II, even at lower energies, although the energy range is difficult to specify. Using equation~\ref{eq:betamin}, and taking a particle with velocity~$\sim800\1{km/s}$, corresponding to a particle near the escape velocity after boosting into the Earth's frame, we can find the relevant energy ranges allowed from the parameter space in figure~\ref{fig:deltamass}. We find that recoil energies as low as~$\sim 15\1{keV}$ are in principle possible at CRESST-II. However, a signal from such high velocities would be highly modulated~\cite{Lang:2010cd}, which would then be potentially clear in an annual cycle of CRESST-II data.

Since the allowed region of parameter space is tuned to suppress the CRESST-II signal while leaving a signal at DAMA, it should be understood that the precise location of the allowed range is sensitive to small changes in form factors, halo models, and other inputs, such as the precise value of the DAMA doping. That said, the existence of a consistent region of parameter space is robust, even if its precise position is not.

\section{Conclusions}
We have re-examined the iDM scenario keeping in mind the (intentional) thallium impurities of the NaI(Tl) crystals used at DAMA. We find this opens up a region of parameter space at large splittings $\delta$. Constraints from the CRESST-II experiment are strong, and the value of $\delta$ in the remaining region of parameter space is tuned at roughly the 5\% level. Nonetheless, this region exists and is a challenge to probe. A highly-modulated signal could arise in CRESST-II, although this is sensitively dependent on the halo model, and difficult to quantify. It is entirely possible that all conventional targets are too light to detect such a scenario, making doped scintillators the target of choice to probe this region of parameter space. In particular, the KIMS experiment, with a doped CsI(Tl) target, should necessarily see the modulation. Variations in the thallium dopant within the DAMA crystals, if controlled, could also test this scenario in future expansions of the DAMA experiment.

\section{Acknowledgements}
The authors thank UC Davis and the HEFTI workshop on light dark matter, where this work was initiated, and Itay Yavin for useful conversations. We are grateful for the availability of the very useful SRIM code. SC is supported by DOE Grant \#DE-FG02-91ER40674. RFL is supported by the NSF grants PHY-0705337 and PHY-0904220. NW is supported by DOE OJI grant \#DE-FG02-06ER41417 and NSF grant \#0947827.


\begin{thebibliography}{48}
\expandafter\ifx\csname natexlab\endcsname\relax\def\natexlab#1{#1}\fi
\expandafter\ifx\csname bibnamefont\endcsname\relax
  \def\bibnamefont#1{#1}\fi
\expandafter\ifx\csname bibfnamefont\endcsname\relax
  \def\bibfnamefont#1{#1}\fi
\expandafter\ifx\csname citenamefont\endcsname\relax
  \def\citenamefont#1{#1}\fi
\expandafter\ifx\csname url\endcsname\relax
  \def\url#1{\texttt{#1}}\fi
\expandafter\ifx\csname urlprefix\endcsname\relax\def\urlprefix{URL }\fi
\providecommand{\bibinfo}[2]{#2}
\providecommand{\eprint}[2][]{\url{#2}}

\bibitem[{\citenamefont{Bernabei et~al.}(2010)}]{Bernabei:2010mq}
\bibinfo{author}{\bibfnamefont{R.}~\bibnamefont{Bernabei}}
  \bibnamefont{et~al.}, \bibinfo{journal}{Eur. Phys. J.}
  \textbf{\bibinfo{volume}{C67}}, \bibinfo{pages}{39} (\bibinfo{year}{2010}).

\bibitem[{\citenamefont{Aprile et~al.}(2010)}]{Aprile:2010um}
\bibinfo{author}{\bibfnamefont{E.}~\bibnamefont{Aprile}} \bibnamefont{et~al.},
  \bibinfo{journal}{Phys. Rev. Lett.} \textbf{\bibinfo{volume}{105}},
  \bibinfo{pages}{131302} (\bibinfo{year}{2010}), \eprint{1005.0380}.

\bibitem[{\citenamefont{Ahmed et~al.}(2010)}]{Ahmed:2009zw}
\bibinfo{author}{\bibfnamefont{Z.}~\bibnamefont{Ahmed}} \bibnamefont{et~al.},
  \bibinfo{journal}{Science} \textbf{\bibinfo{volume}{327}},
  \bibinfo{pages}{1619} (\bibinfo{year}{2010}).

\bibitem[{\citenamefont{Angle et~al.}(2008)}]{Angle:2007uj}
\bibinfo{author}{\bibfnamefont{J.}~\bibnamefont{Angle}} \bibnamefont{et~al.},
  \bibinfo{journal}{Phys. Rev. Lett.} \textbf{\bibinfo{volume}{100}},
  \bibinfo{pages}{021303} (\bibinfo{year}{2008}).

\bibitem[{\citenamefont{Angloher et~al.}(2009)}]{Angloher:2008jj}
\bibinfo{author}{\bibfnamefont{G.}~\bibnamefont{Angloher}}
  \bibnamefont{et~al.}, \bibinfo{journal}{Astropart. Phys.}
  \textbf{\bibinfo{volume}{31}}, \bibinfo{pages}{270} (\bibinfo{year}{2009}).

\bibitem[{\citenamefont{Lee et~al.}(2007)}]{Lee.:2007qn}
\bibinfo{author}{\bibfnamefont{H.~S.} \bibnamefont{Lee}} \bibnamefont{et~al.},
  \bibinfo{journal}{Phys. Rev. Lett.} \textbf{\bibinfo{volume}{99}},
  \bibinfo{pages}{091301} (\bibinfo{year}{2007}).

\bibitem[{\citenamefont{Lebedenko et~al.}(2009)}]{Lebedenko:2008gb}
\bibinfo{author}{\bibfnamefont{V.~N.} \bibnamefont{Lebedenko}}
  \bibnamefont{et~al.}, \bibinfo{journal}{Phys. Rev.}
  \textbf{\bibinfo{volume}{D80}}, \bibinfo{pages}{052010}
  (\bibinfo{year}{2009}).

\bibitem[{\citenamefont{Behnke et~al.}(2008)}]{Behnke:2008zza}
\bibinfo{author}{\bibfnamefont{E.}~\bibnamefont{Behnke}} \bibnamefont{et~al.},
  \bibinfo{journal}{Science} \textbf{\bibinfo{volume}{319}},
  \bibinfo{pages}{933} (\bibinfo{year}{2008}).

\bibitem[{\citenamefont{Bottino et~al.}(2004)\citenamefont{Bottino, Donato,
  Fornengo, and Scopel}}]{Bottino:2003cz}
\bibinfo{author}{\bibfnamefont{A.}~\bibnamefont{Bottino}},
  \bibinfo{author}{\bibfnamefont{F.}~\bibnamefont{Donato}},
  \bibinfo{author}{\bibfnamefont{N.}~\bibnamefont{Fornengo}}, \bibnamefont{and}
  \bibinfo{author}{\bibfnamefont{S.}~\bibnamefont{Scopel}},
  \bibinfo{journal}{Phys. Rev.} \textbf{\bibinfo{volume}{D69}},
  \bibinfo{pages}{037302} (\bibinfo{year}{2004}).

\bibitem[{\citenamefont{Belli et~al.}(2000)}]{Belli:1999nz}
\bibinfo{author}{\bibfnamefont{P.}~\bibnamefont{Belli}} \bibnamefont{et~al.},
  \bibinfo{journal}{Phys. Rev.} \textbf{\bibinfo{volume}{D61}},
  \bibinfo{pages}{023512} (\bibinfo{year}{2000}).

\bibitem[{\citenamefont{Petriello and Zurek}(2008)}]{Petriello:2008jj}
\bibinfo{author}{\bibfnamefont{F.}~\bibnamefont{Petriello}} \bibnamefont{and}
  \bibinfo{author}{\bibfnamefont{K.~M.} \bibnamefont{Zurek}},
  \bibinfo{journal}{JHEP} \textbf{\bibinfo{volume}{09}}, \bibinfo{pages}{047}
  (\bibinfo{year}{2008}).

\bibitem[{\citenamefont{Chang et~al.}(2009{\natexlab{a}})\citenamefont{Chang,
  Pierce, and Weiner}}]{Chang:2008xa}
\bibinfo{author}{\bibfnamefont{S.}~\bibnamefont{Chang}},
  \bibinfo{author}{\bibfnamefont{A.}~\bibnamefont{Pierce}}, \bibnamefont{and}
  \bibinfo{author}{\bibfnamefont{N.}~\bibnamefont{Weiner}},
  \bibinfo{journal}{Phys. Rev.} \textbf{\bibinfo{volume}{D79}},
  \bibinfo{pages}{115011} (\bibinfo{year}{2009}{\natexlab{a}}).

\bibitem[{\citenamefont{Savage et~al.}(2009)\citenamefont{Savage, Gelmini,
  Gondolo, and Freese}}]{Savage:2008er}
\bibinfo{author}{\bibfnamefont{C.}~\bibnamefont{Savage}},
  \bibinfo{author}{\bibfnamefont{G.}~\bibnamefont{Gelmini}},
  \bibinfo{author}{\bibfnamefont{P.}~\bibnamefont{Gondolo}}, \bibnamefont{and}
  \bibinfo{author}{\bibfnamefont{K.}~\bibnamefont{Freese}},
  \bibinfo{journal}{JCAP} \textbf{\bibinfo{volume}{0904}}, \bibinfo{pages}{010}
  (\bibinfo{year}{2009}).

\bibitem[{\citenamefont{Hooper et~al.}(2010)\citenamefont{Hooper, Collar, Hall,
  and McKinsey}}]{Hooper:2010uy}
\bibinfo{author}{\bibfnamefont{D.}~\bibnamefont{Hooper}},
  \bibinfo{author}{\bibfnamefont{J.~I.} \bibnamefont{Collar}},
  \bibinfo{author}{\bibfnamefont{J.}~\bibnamefont{Hall}}, \bibnamefont{and}
  \bibinfo{author}{\bibfnamefont{D.}~\bibnamefont{McKinsey}}
  (\bibinfo{year}{2010}), \eprint{1007.1005}.

\bibitem[{\citenamefont{Ullio et~al.}(2001)\citenamefont{Ullio, Kamionkowski,
  and Vogel}}]{Ullio:2000bv}
\bibinfo{author}{\bibfnamefont{P.}~\bibnamefont{Ullio}},
  \bibinfo{author}{\bibfnamefont{M.}~\bibnamefont{Kamionkowski}},
  \bibnamefont{and} \bibinfo{author}{\bibfnamefont{P.}~\bibnamefont{Vogel}},
  \bibinfo{journal}{JHEP} \textbf{\bibinfo{volume}{07}}, \bibinfo{pages}{044}
  (\bibinfo{year}{2001}).

\bibitem[{\citenamefont{Belli et~al.}(2002)\citenamefont{Belli, Cerulli,
  Fornengo, and Scopel}}]{Belli:2002yt}
\bibinfo{author}{\bibfnamefont{P.}~\bibnamefont{Belli}},
  \bibinfo{author}{\bibfnamefont{R.}~\bibnamefont{Cerulli}},
  \bibinfo{author}{\bibfnamefont{N.}~\bibnamefont{Fornengo}}, \bibnamefont{and}
  \bibinfo{author}{\bibfnamefont{S.}~\bibnamefont{Scopel}},
  \bibinfo{journal}{Phys. Rev.} \textbf{\bibinfo{volume}{D66}},
  \bibinfo{pages}{043503} (\bibinfo{year}{2002}).

\bibitem[{\citenamefont{Savage et~al.}(2004)\citenamefont{Savage, Gondolo, and
  Freese}}]{Savage:2004fn}
\bibinfo{author}{\bibfnamefont{C.}~\bibnamefont{Savage}},
  \bibinfo{author}{\bibfnamefont{P.}~\bibnamefont{Gondolo}}, \bibnamefont{and}
  \bibinfo{author}{\bibfnamefont{K.}~\bibnamefont{Freese}},
  \bibinfo{journal}{Phys. Rev.} \textbf{\bibinfo{volume}{D70}},
  \bibinfo{pages}{123513} (\bibinfo{year}{2004}).

\bibitem[{\citenamefont{Fairbairn and Schwetz}(2009)}]{Fairbairn:2008gz}
\bibinfo{author}{\bibfnamefont{M.}~\bibnamefont{Fairbairn}} \bibnamefont{and}
  \bibinfo{author}{\bibfnamefont{T.}~\bibnamefont{Schwetz}},
  \bibinfo{journal}{JCAP} \textbf{\bibinfo{volume}{0901}}, \bibinfo{pages}{037}
  (\bibinfo{year}{2009}).

\bibitem[{\citenamefont{Foot}(2008)}]{Foot:2008nw}
\bibinfo{author}{\bibfnamefont{R.}~\bibnamefont{Foot}}, \bibinfo{journal}{Phys.
  Rev.} \textbf{\bibinfo{volume}{D78}}, \bibinfo{pages}{043529}
  (\bibinfo{year}{2008}).

\bibitem[{\citenamefont{Feldstein et~al.}(2010)\citenamefont{Feldstein,
  Fitzpatrick, and Katz}}]{Feldstein:2009tr}
\bibinfo{author}{\bibfnamefont{B.}~\bibnamefont{Feldstein}},
  \bibinfo{author}{\bibfnamefont{A.~L.} \bibnamefont{Fitzpatrick}},
  \bibnamefont{and} \bibinfo{author}{\bibfnamefont{E.}~\bibnamefont{Katz}},
  \bibinfo{journal}{JCAP} \textbf{\bibinfo{volume}{1001}}, \bibinfo{pages}{020}
  (\bibinfo{year}{2010}).

\bibitem[{\citenamefont{Chang et~al.}(2010{\natexlab{a}})\citenamefont{Chang,
  Pierce, and Weiner}}]{Chang:2009yt}
\bibinfo{author}{\bibfnamefont{S.}~\bibnamefont{Chang}},
  \bibinfo{author}{\bibfnamefont{A.}~\bibnamefont{Pierce}}, \bibnamefont{and}
  \bibinfo{author}{\bibfnamefont{N.}~\bibnamefont{Weiner}},
  \bibinfo{journal}{JCAP} \textbf{\bibinfo{volume}{1001}}, \bibinfo{pages}{006}
  (\bibinfo{year}{2010}{\natexlab{a}}).

\bibitem[{\citenamefont{Tucker-Smith and Weiner}(2001)}]{TuckerSmith:2001hy}
\bibinfo{author}{\bibfnamefont{D.}~\bibnamefont{Tucker-Smith}}
  \bibnamefont{and} \bibinfo{author}{\bibfnamefont{N.}~\bibnamefont{Weiner}},
  \bibinfo{journal}{Phys. Rev.} \textbf{\bibinfo{volume}{D64}},
  \bibinfo{pages}{043502} (\bibinfo{year}{2001}).

\bibitem[{\citenamefont{Tucker-Smith and Weiner}(2005)}]{TuckerSmith:2004jv}
\bibinfo{author}{\bibfnamefont{D.}~\bibnamefont{Tucker-Smith}}
  \bibnamefont{and} \bibinfo{author}{\bibfnamefont{N.}~\bibnamefont{Weiner}},
  \bibinfo{journal}{Phys. Rev.} \textbf{\bibinfo{volume}{D72}},
  \bibinfo{pages}{063509} (\bibinfo{year}{2005}).

\bibitem[{\citenamefont{Chang et~al.}(2009{\natexlab{b}})\citenamefont{Chang,
  Kribs, Tucker-Smith, and Weiner}}]{Chang:2008gd}
\bibinfo{author}{\bibfnamefont{S.}~\bibnamefont{Chang}},
  \bibinfo{author}{\bibfnamefont{G.~D.} \bibnamefont{Kribs}},
  \bibinfo{author}{\bibfnamefont{D.}~\bibnamefont{Tucker-Smith}},
  \bibnamefont{and} \bibinfo{author}{\bibfnamefont{N.}~\bibnamefont{Weiner}},
  \bibinfo{journal}{Phys. Rev.} \textbf{\bibinfo{volume}{D79}},
  \bibinfo{pages}{043513} (\bibinfo{year}{2009}{\natexlab{b}}).

\bibitem[{\citenamefont{March-Russell et~al.}(2009)\citenamefont{March-Russell,
  McCabe, and McCullough}}]{MarchRussell:2008dy}
\bibinfo{author}{\bibfnamefont{J.}~\bibnamefont{March-Russell}},
  \bibinfo{author}{\bibfnamefont{C.}~\bibnamefont{McCabe}}, \bibnamefont{and}
  \bibinfo{author}{\bibfnamefont{M.}~\bibnamefont{McCullough}},
  \bibinfo{journal}{JHEP} \textbf{\bibinfo{volume}{05}}, \bibinfo{pages}{071}
  (\bibinfo{year}{2009}), \eprint{0812.1931}.

\bibitem[{\citenamefont{Kuhlen et~al.}(2010)}]{Kuhlen:2009vh}
\bibinfo{author}{\bibfnamefont{M.}~\bibnamefont{Kuhlen}} \bibnamefont{et~al.},
  \bibinfo{journal}{JCAP} \textbf{\bibinfo{volume}{1002}}, \bibinfo{pages}{030}
  (\bibinfo{year}{2010}).

\bibitem[{\citenamefont{Kopp et~al.}(2010)\citenamefont{Kopp, Schwetz, and
  Zupan}}]{Kopp:2009qt}
\bibinfo{author}{\bibfnamefont{J.}~\bibnamefont{Kopp}},
  \bibinfo{author}{\bibfnamefont{T.}~\bibnamefont{Schwetz}}, \bibnamefont{and}
  \bibinfo{author}{\bibfnamefont{J.}~\bibnamefont{Zupan}},
  \bibinfo{journal}{JCAP} \textbf{\bibinfo{volume}{1002}}, \bibinfo{pages}{014}
  (\bibinfo{year}{2010}).

\bibitem[{\citenamefont{Schmidt-Hoberg and
  Winkler}(2009)}]{SchmidtHoberg:2009gn}
\bibinfo{author}{\bibfnamefont{K.}~\bibnamefont{Schmidt-Hoberg}}
  \bibnamefont{and} \bibinfo{author}{\bibfnamefont{M.~W.}
  \bibnamefont{Winkler}}, \bibinfo{journal}{JCAP}
  \textbf{\bibinfo{volume}{0909}}, \bibinfo{pages}{010} (\bibinfo{year}{2009}).

\bibitem[{\citenamefont{Akimov et~al.}(2010)}]{Akimov:2010vk}
\bibinfo{author}{\bibfnamefont{D.~Y.} \bibnamefont{Akimov}}
  \bibnamefont{et~al.}, \bibinfo{journal}{Phys. Lett.}
  \textbf{\bibinfo{volume}{B692}}, \bibinfo{pages}{180} (\bibinfo{year}{2010}),
  \eprint{1003.5626}.

\bibitem[{\citenamefont{Lang and Weiner}(2010)}]{Lang:2010cd}
\bibinfo{author}{\bibfnamefont{R.~F.} \bibnamefont{Lang}} \bibnamefont{and}
  \bibinfo{author}{\bibfnamefont{N.}~\bibnamefont{Weiner}},
  \bibinfo{journal}{JCAP} \textbf{\bibinfo{volume}{1006}}, \bibinfo{pages}{032}
  (\bibinfo{year}{2010}).

\bibitem[{\citenamefont{Alves et~al.}(2010)\citenamefont{Alves, Lisanti, and
  Wacker}}]{Alves:2010pt}
\bibinfo{author}{\bibfnamefont{D.~S.~M.} \bibnamefont{Alves}},
  \bibinfo{author}{\bibfnamefont{M.}~\bibnamefont{Lisanti}}, \bibnamefont{and}
  \bibinfo{author}{\bibfnamefont{J.~G.} \bibnamefont{Wacker}},
  \bibinfo{journal}{Phys. Rev.} \textbf{\bibinfo{volume}{D82}},
  \bibinfo{pages}{031901} (\bibinfo{year}{2010}), \eprint{1005.5421}.

\bibitem[{\citenamefont{Bernabei et~al.}(2008)}]{Bernabei:2008yh}
\bibinfo{author}{\bibfnamefont{R.}~\bibnamefont{Bernabei}}
  \bibnamefont{et~al.}, \bibinfo{journal}{Nucl. Instrum. Meth.}
  \textbf{\bibinfo{volume}{A592}}, \bibinfo{pages}{297} (\bibinfo{year}{2008}).

\bibitem[{\citenamefont{Helm}(1956)}]{Helm:1956zz}
\bibinfo{author}{\bibfnamefont{R.~H.} \bibnamefont{Helm}},
  \bibinfo{journal}{Phys. Rev.} \textbf{\bibinfo{volume}{104}},
  \bibinfo{pages}{1466} (\bibinfo{year}{1956}).

\bibitem[{\citenamefont{Arkani-Hamed et~al.}(2009)\citenamefont{Arkani-Hamed,
  Finkbeiner, Slatyer, and Weiner}}]{ArkaniHamed:2008qn}
\bibinfo{author}{\bibfnamefont{N.}~\bibnamefont{Arkani-Hamed}},
  \bibinfo{author}{\bibfnamefont{D.~P.} \bibnamefont{Finkbeiner}},
  \bibinfo{author}{\bibfnamefont{T.~R.} \bibnamefont{Slatyer}},
  \bibnamefont{and} \bibinfo{author}{\bibfnamefont{N.}~\bibnamefont{Weiner}},
  \bibinfo{journal}{Phys. Rev.} \textbf{\bibinfo{volume}{D79}},
  \bibinfo{pages}{015014} (\bibinfo{year}{2009}).

\bibitem[{\citenamefont{Baumgart et~al.}(2009)\citenamefont{Baumgart, Cheung,
  Ruderman, Wang, and Yavin}}]{Baumgart:2009tn}
\bibinfo{author}{\bibfnamefont{M.}~\bibnamefont{Baumgart}},
  \bibinfo{author}{\bibfnamefont{C.}~\bibnamefont{Cheung}},
  \bibinfo{author}{\bibfnamefont{J.~T.} \bibnamefont{Ruderman}},
  \bibinfo{author}{\bibfnamefont{L.-T.} \bibnamefont{Wang}}, \bibnamefont{and}
  \bibinfo{author}{\bibfnamefont{I.}~\bibnamefont{Yavin}},
  \bibinfo{journal}{JHEP} \textbf{\bibinfo{volume}{04}}, \bibinfo{pages}{014}
  (\bibinfo{year}{2009}).

\bibitem[{\citenamefont{Cheung et~al.}(2009)\citenamefont{Cheung, Ruderman,
  Wang, and Yavin}}]{Cheung:2009qd}
\bibinfo{author}{\bibfnamefont{C.}~\bibnamefont{Cheung}},
  \bibinfo{author}{\bibfnamefont{J.~T.} \bibnamefont{Ruderman}},
  \bibinfo{author}{\bibfnamefont{L.-T.} \bibnamefont{Wang}}, \bibnamefont{and}
  \bibinfo{author}{\bibfnamefont{I.}~\bibnamefont{Yavin}},
  \bibinfo{journal}{Phys. Rev.} \textbf{\bibinfo{volume}{D80}},
  \bibinfo{pages}{035008} (\bibinfo{year}{2009}).

\bibitem[{\citenamefont{Morrissey et~al.}(2009)\citenamefont{Morrissey, Poland,
  and Zurek}}]{Morrissey:2009ur}
\bibinfo{author}{\bibfnamefont{D.~E.} \bibnamefont{Morrissey}},
  \bibinfo{author}{\bibfnamefont{D.}~\bibnamefont{Poland}}, \bibnamefont{and}
  \bibinfo{author}{\bibfnamefont{K.~M.} \bibnamefont{Zurek}},
  \bibinfo{journal}{JHEP} \textbf{\bibinfo{volume}{07}}, \bibinfo{pages}{050}
  (\bibinfo{year}{2009}).

\bibitem[{\citenamefont{Bjorken et~al.}(2009)\citenamefont{Bjorken, Essig,
  Schuster, and Toro}}]{Bjorken:2009mm}
\bibinfo{author}{\bibfnamefont{J.~D.} \bibnamefont{Bjorken}},
  \bibinfo{author}{\bibfnamefont{R.}~\bibnamefont{Essig}},
  \bibinfo{author}{\bibfnamefont{P.}~\bibnamefont{Schuster}}, \bibnamefont{and}
  \bibinfo{author}{\bibfnamefont{N.}~\bibnamefont{Toro}},
  \bibinfo{journal}{Phys. Rev.} \textbf{\bibinfo{volume}{D80}},
  \bibinfo{pages}{075018} (\bibinfo{year}{2009}).

\bibitem[{\citenamefont{Bavykina et~al.}(2007)}]{Bavykina:2007ze}
\bibinfo{author}{\bibfnamefont{I.}~\bibnamefont{Bavykina}}
  \bibnamefont{et~al.}, \bibinfo{journal}{Astropart. Phys.}
  \textbf{\bibinfo{volume}{28}}, \bibinfo{pages}{489} (\bibinfo{year}{2007}).

\bibitem[{\citenamefont{{J.~Ziegler}}({2010})}]{srim}
\bibinfo{author}{\bibnamefont{{J.~Ziegler}}}, \emph{\bibinfo{title}{{The
  Stopping and Range of Ions in Matter}}}, \bibinfo{howpublished}{{software
  available from \url{www.srim.org}}} (\bibinfo{year}{{2010}}).

\bibitem[{\citenamefont{Fushimi et~al.}(1993)}]{Fushimi:1993nq}
\bibinfo{author}{\bibfnamefont{K.}~\bibnamefont{Fushimi}} \bibnamefont{et~al.},
  \bibinfo{journal}{Phys. Rev.} \textbf{\bibinfo{volume}{C47}},
  \bibinfo{pages}{425} (\bibinfo{year}{1993}).

\bibitem[{\citenamefont{Tovey et~al.}(1998)}]{Tovey:1998ex}
\bibinfo{author}{\bibfnamefont{D.~R.} \bibnamefont{Tovey}}
  \bibnamefont{et~al.}, \bibinfo{journal}{Phys. Lett.}
  \textbf{\bibinfo{volume}{B433}}, \bibinfo{pages}{150} (\bibinfo{year}{1998}).

\bibitem[{\citenamefont{Gerbier et~al.}(1999)}]{Gerbier:1998dm}
\bibinfo{author}{\bibfnamefont{G.}~\bibnamefont{Gerbier}} \bibnamefont{et~al.},
  \bibinfo{journal}{Astropart. Phys.} \textbf{\bibinfo{volume}{11}},
  \bibinfo{pages}{287} (\bibinfo{year}{1999}).

\bibitem[{\citenamefont{Lang and Seidel}(2009)}]{Lang:2009ge}
\bibinfo{author}{\bibfnamefont{R.~F.} \bibnamefont{Lang}} \bibnamefont{and}
  \bibinfo{author}{\bibfnamefont{W.}~\bibnamefont{Seidel}},
  \bibinfo{journal}{New J. Phys.} \textbf{\bibinfo{volume}{11}},
  \bibinfo{pages}{105017} (\bibinfo{year}{2009}).

\bibitem[{\citenamefont{Lang et~al.}(2010)}]{Lang:2009wb}
\bibinfo{author}{\bibfnamefont{R.~F.} \bibnamefont{Lang}} \bibnamefont{et~al.},
  \bibinfo{journal}{Astropart. Phys.} \textbf{\bibinfo{volume}{32}},
  \bibinfo{pages}{318} (\bibinfo{year}{2010}).

\bibitem[{\citenamefont{Cozzini et~al.}(2004)}]{Cozzini:2004vd}
\bibinfo{author}{\bibfnamefont{C.}~\bibnamefont{Cozzini}} \bibnamefont{et~al.},
  \bibinfo{journal}{Phys. Rev.} \textbf{\bibinfo{volume}{C70}},
  \bibinfo{pages}{064606} (\bibinfo{year}{2004}).

\bibitem[{\citenamefont{Park et~al.}(2002)}]{Park:2002jr}
\bibinfo{author}{\bibfnamefont{H.}~\bibnamefont{Park}} \bibnamefont{et~al.},
  \bibinfo{journal}{Nucl. Instrum. Meth.} \textbf{\bibinfo{volume}{A491}},
  \bibinfo{pages}{460} (\bibinfo{year}{2002}).

\bibitem[{\citenamefont{Chang et~al.}(2010{\natexlab{b}})\citenamefont{Chang,
  Weiner, and Yavin}}]{Chang:2010en}
\bibinfo{author}{\bibfnamefont{S.}~\bibnamefont{Chang}},
  \bibinfo{author}{\bibfnamefont{N.}~\bibnamefont{Weiner}}, \bibnamefont{and}
  \bibinfo{author}{\bibfnamefont{I.}~\bibnamefont{Yavin}}
  (\bibinfo{year}{2010}{\natexlab{b}}), \eprint{1007.4200}.

\end{thebibliography}

\end{document}